# Report on Spring8 Test

G. Gosta[1,2]

1 University of Milano, Department of Physics, Via Celoria 16, 20133 Milano, Italy

2 INFN Section of Milano, Via Celoria 16, 20133 Milano, Italy

## Abstract

*γ-rays in the energy range 6 – 38 MeV were produced and sent into two large volume $LaBr_3$:Ce crystals (3.5"x8") at the NewSUBARU facility, with the goal of investigating the response function of the detectors. By comparing the experimental spectra of the two detectors we deduced the linearity of the system, separately of the two crystals and of the two PMT + VD associated. Moreover, Monte Carlo simulations were performed in order to reproduce the experimental spectra. The photopeak and interaction efficiencies both in case of a collimated beam and an isotropic source were also evaluated.*

## I. Introduction

The study of the collective properties of nuclei, like the Giant Dipole Resonance (built on the ground state [1, 2] or on an excited state [3, 4]) or like the Pigmy Dipole Resonance [5, 6], usually implies the measurement of a continuum of high energy γ-rays (5 < $E_γ$ < 30 MeV) [3, 4] and requires the knowledge of the detector response function.

Standard calibration sources can provide γ-rays of at most 9 MeV [11]. Therefore, in order to extend the detector calibration to energies higher than 9 MeV, one should use in-beam nuclear reactions which produce high energy monochromatic γ-rays (see for example Ref. [12]) to be used for calibration and/or extrapolate the calibration supposing a linear or a specific behavior. However, the use of in-beam measurements is often impossible for practical reasons, and the extrapolation is susceptible to large errors. The response function is generally calculated using Monte Carlo simulations (i.e. Geant4 [7]) supposing an ideal behavior of the detector.

In the past years, GDR studies were performed using large volume NaI [8] or $BaF_2$ [3] scintillators. Recently, the development of the $LaBr_3$:Ce material as a scintillator was a major improvement in the field of scintillator detectors for the measurement of high energy gamma rays [9].

What makes the $LaBr_3$:Ce an extremely good scintillator crystal is the origin of nonlinear effects when high energy γ-rays need to be measured. A high light yield (63000 ph/MeV) and a very short decay time of the scintillation light induce a large current inside the Phototube (PMT) [17].

The response function and the linearity at very low energies ($E_γ$ < 100 keV) was already accurately studied for example by Alekhin et al. [13] and a crystal non-linearity up to 15% was measured. In several papers (see for example [14, 15] ) the linearity of large $LaBr_3$:Ce was studied for γ-rays up to ≈ 10 MeV using calibration sources.

Here we report the measurement of the response function of two scintillator detectors (each consists of a 3.5"x8" large volume LaBr$_3$:Ce crystal) to quasi-monochromatic γ-rays with energy from 6 to 38 MeV produced in the NewSUBARU facility [18]. Each crystal was coupled to two PMT's so that for each beam energy four spectra were recorded. A more complete report of these results can be found in ref [26]

We investigated the behavior of the crystal, of the PMT together with the Voltage Divider and the electronics used to handle the signals in a separate way, so that we could identify the origin of possible non-linearity effects.

It turns out that the two crystals show the same behavior, while a different non-linearity is observed for the two PMT's. A non-linearity curve was deduced for each PMT by performing Monte Carlo simulations.

To perform the measurement, we used two LaBr$_3$(Ce) detectors, two PMTs and two VDs:

- The two large volume cylindrical LaBr$_3$:Ce crystals (3.5"x8") are produced by St. Gobain and both crystals are sealed in an aluminum capsule with a glass window for the PMT, as the LaBr$_3$:Ce material is strongly hygroscopic.
- The crystal with serial number L824 will be indicated as crystal 1 (shortened C1) while the one with serial number K604 will be indicated as crystal 2 (C2). In the reference sheet provided by the company the declared crystal energy resolution was 3.1% for both crystals.
- The used PMT's are model R10233-100SEL from Hamamatsu with a serial number ZE5555 and ZE5559. The PMT's were coupled to the LaBr$_3$:Ce crystals with BC-630 optical grease. The R10233-100 specifications as reported in the PMT reference sheet are listed in table 1.
- The Voltage Dividers (VD), identical in both detectors, have an active design and were especially developed for large volume LaBr$_3$:Ce crystals at the University of Milano [10]. The VD was tuned experimentally to preserve the intrinsically good energy and timing properties of the crystals while obtaining at the same time a relatively homogeneous energy response linearity among the various PMT parts. The performances of these VD's were already discussed in Ref. [10].
- For simplicity, we will label the two systems PMT+VD as P55 and P59 for the ZE5555 and ZE5559 PMT's, respectively.
- The detector anode signal was sent to a spectroscopic amplifier especially designed by the INFN section of Milano (LABPRO) [19, 20, 21].

| PMT S/N | Cathode Lum. Sens. µA/lm | Anode Lum. Sens. A/lm | Anode Dark current nA | Cathode Blue Sens. Index |
|---|---|---|---|---|
| ZE5555 | 143.0 | 20.1 | 3.9 | 14.80 |
| ZE5559 | 150.0 | 7.1 | 0.44 | 15.60 |

Table 1: *The R10233-100 specifications as reported in the PMT reference sheet are listed. They refer to measurements performed at voltage of 1000 V using the standard voltage distribution ratio listed in the HAMAMATSU photomultiplier catalog.*

A measurement of the linear response of LABPRO was performed using a trapezoidal pulse with a rise time of 30 ns, which mimic the LaBr$_3$:Ce anode pulse. The signals were provided to LABPRO by an Agilent 33220A pulser. The output signal of LABPRO was split and sent to two Ortec ADCAM 926 MCA. The system Pulser + LABPRO + MCA was found to be linear within 0.3% up to 5.5 V.

With this electronic set up, we measured the energy resolution at 662 keV using a $^{137}$Cs source for different values of the amplifier gain. We used crystal 1 coupled to P55 and crystal 2 coupled to P59. We obtained that the resolution is affected only slightly by the amplification of the signal. As this kind of measure implies a wide dynamic range, we used the lowest possible gain, for which we obtained an energy resolution of about 20.5 keV for C1 and 20.7 keV for C2.

The measurement was performed in the GACKO experimental hutch of the NewSUBARU synchrotron radiation facility located in the Spring8 site [18]. The facility provides a quasi-monochromatic γ beam using the Laser Compton Scattering (LCS) mechanism. Namely, the high energy γ-rays are produced through a collision between high energy electrons (0.5-1.5 GeV), accumulated in the NewSUBARU storage ring, and 1064 nm CW photons produced by a Nd:YVO$_4$ laser [18] which has a maximum power of 35 watt. The so produced γ-rays pass through two collimators.

II. *Incident photon distribution*:

The incident γ-ray energy distributions were obtained by the ELI-NP Group performing GEANT4 simulations of the response functions of the LaBr$_3$:Ce detector to the LCS gamma-ray beam, involving the kinematics of LCS and collimation geometry as explained in detail in Ref. [1].

The maximum incident photon spectra hitting the LaBr$_3$:Ce detector is displayed in table 2 togheter with the nominal electron-beam energies in the NewSUBARU ring. It is important to remember that the beam is less and less monochromatic as the beam energy increases. In particular, while the distribution is quite sharp on the high-energy side of the maximum value, it becomes broader on the low-energy side with increasing beam energy. In this report, we will define as beam energy the maximum energy of the γ-beam, keeping in mind that for high energy only 10-20% of the beam γ-rays have energies close to the maximum energy.

In order to maintain the count rate below few KHz and to avoid pile-up, a 5 cm thick lead brick was inserted in the beam line trajectory about two meters before the detectors.

| Electron Energy [GeV] | 0.575 | 0.704 | 0.849 | 0.914 | 0.974 | 1.087 | 1.162 | 1.273 | 1.380 | 1.460 |
|---|---|---|---|---|---|---|---|---|---|---|
| γ-ray Energy [MeV] | 6 | 9 | 13 | 15 | 17 | 21 | 24 | 29 | 34 | 38 |

Table 2: *The electron energies in the NewSUBARU ring and the corresponding maximum energies $E_{max}$ produced in the LCS hitting detector are listed.*

The electron beam energy was calibrated with an accuracy of $10^{-5}$ in the nominal electron energy range of 550 – 1460 MeV [22,23]. Thus, the maximum energy of the LCS gamma-ray beam was determined by the calibrated electron energy and laser photon energy.

### III. *Response function*:

The two LaBr$_3$:Ce detectors were placed horizontally on a trolley (see Fig.1) allowing to center the detectors along the beam line for each measurement.

Ten beam energies were used ranging from 6 to 38 MeV. For each beam energy, four spectra were recorded, since each crystal was coupled to both PMT's. For simplicity, we will label each detector set up with the crystal number and the PMT label as C1-P55, C1-P59, C2-P55 and C2-P59.

In order to be able to subtract the crystal internal radiation contribution to the spectra, the LABPRO output signal was sent to two ADC AMPTEK Mod. MCA8000D. Each ADC was gated by a signal 50ms wide, which activated the laser in one case (Gate-ON) and deactivated it in the other case (Gate-OFF), so that the ADC with the Gate-ON was active in-beam while the ADC with the Gate-OFF was active off-beam.

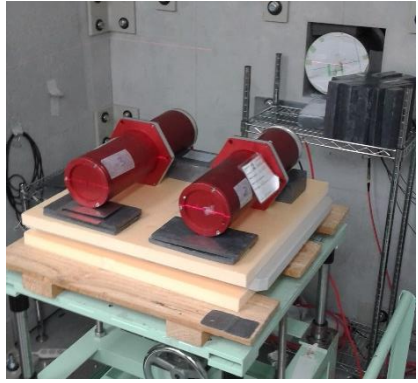

**Fig. 1.** *A picture of the experimental setup. The γ-ray beam direction is indicated by the black arrow. The two LaBr₃:Ce detectors can be moved left-right to intercept the beam*

The count rate of the LaBr$_3$:Ce did not exceed 6 KHz for both ADCs. The high voltage powering the PMT's, given by a CAEN module N1470AL (S/N 0768), was set such that the two anode signals were both approximately 30 mV high for a deposition in the crystal of 661.6 keV. In particular, P55 was powered at 810 V while P59 was powered at 950 V, for all measurements. The electronic noise, measured at the spectroscopic amplifier output, was of the order of 3 mV.

In order to monitor possible signal amplitude variations during the measurements, calibrations were performed before and after each run using standard sealed $^{137}$Cs and $^{60}$Co sources. The analysis of the calibration spectra did not evidence changes in gain.

## IV. Crystals response

Following the comparison between the spectra recorded with C1 and C2 coupled to the same PMT+VD (P55 or P59) it is possible to verify that the two crystals have the same response to high energy γ-rays. In fact, the spectra are very similar in shape for all incident energies. In figure 2, the overlap of the spectra for the two crystals and the same PMT (in this case P59) at beam energies of 6, 15 and 29 MeV are shown.

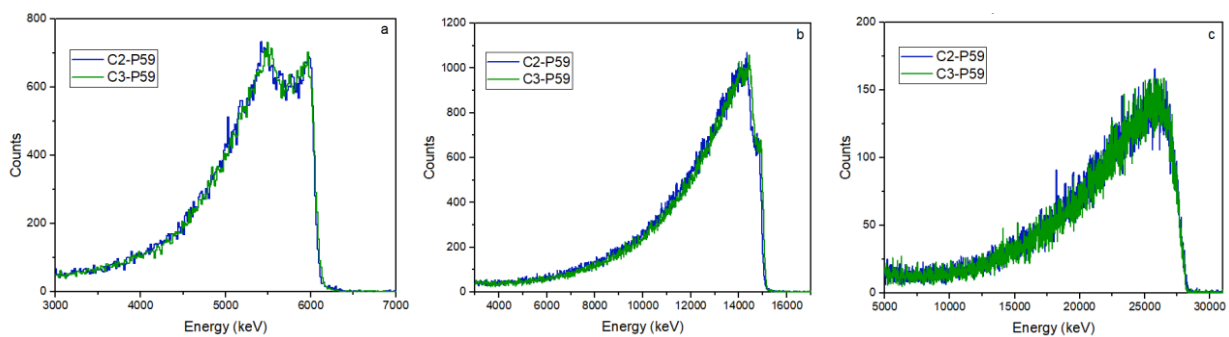

Fig. 2. *The energy spectra measured by C2-P59 (black line) and C1-P59 (red line) are compared for incident γ-rays energy of 6 (a), 15 (b) and 29 MeV(c) (color online).*

At energies higher than the full energy peak, the spectra consist of a flat background not higher than 5 counts. After background subtraction, the spectra end sharply at a channel which we will define as endpoint. For a numerical comparison, we calculated the differences between the endpoints measured with the two crystals coupled to the same PMT+VD.

The results, plotted in Fig. 3 for both the PMTs and beam energies, indicate that the behavior of two crystals is the same within the error bars. We will then assume that the intrinsic response of the two LaBr$_3$:Ce crystals, having the same shape and dimension, does not depend on the particular crystal. The effect evident in ref. [16] was not observed within the error bars.

The size of the error bars could be significantly reduced using an incident radiation with an energy width smaller than the energy resolution of the detectors. These γ beams are not available now but they will be produced, for example, by the new ELI-NP facility [24] in Bucharest (Romania).

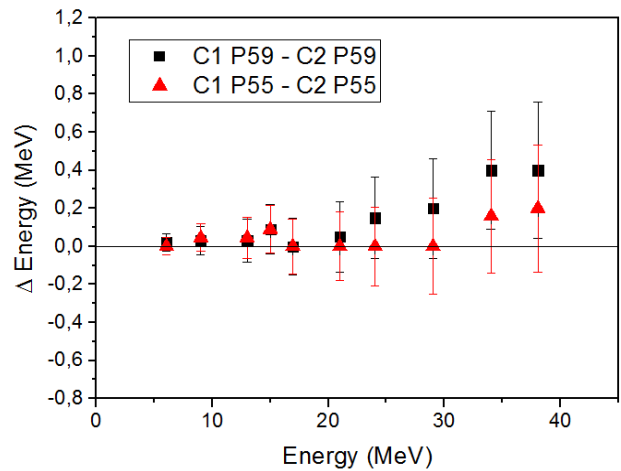

Fig. 3. *The measured energy differences between the position of the endpoint measured in C1 and C2 using the same PMT (indicated by filled circle for the P55 and filled square for the P59) are shown. The error bars were estimated propagating the energy calibration error (color online).*

### V. PMT's + Voltage Divider response

In order to obtain the effect of non-linearity caused by the PMTs, we can compare the spectra recorded by the same crystal coupled to the two different PMTs.

In Fig. 4, the spectra recorded with C2 coupled to P55 and P59 at beam energies of 6, 15 and 29 MeV are compared. The endpoint the two spectra show a difference in the measured energy that increase as the incident γ-ray energy increases. This indicates that the two PMT's have a different non-linearity curve, even though the PMT are of the same type (R10233) and serial number (ZE5555 and ZE5559) is similar.

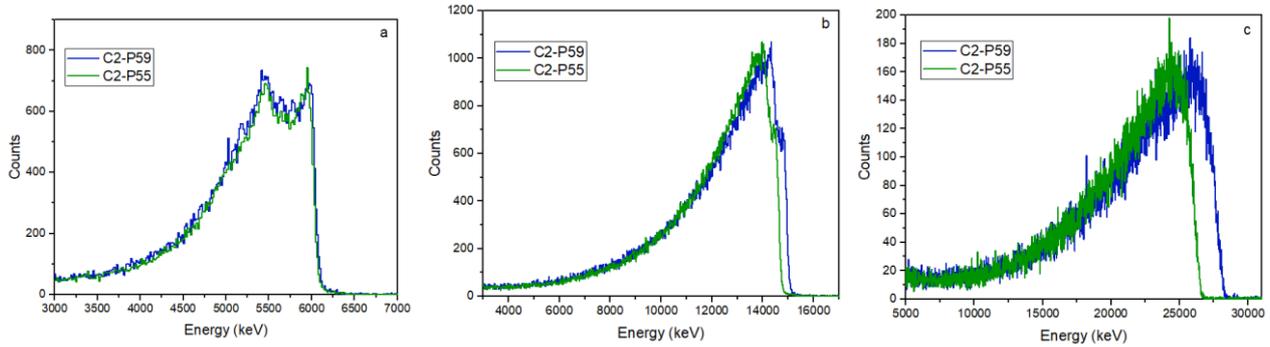

Fig. 4. *The energy spectra measured by C2 P59 (black line) and C2 P55 (red line) are compared at incident γ-rays energies of 6 (a), 15 (b) and 29 MeV (c) (color online).*

However, in order to extract an absolute non-linearity curve for both the PMT's we need to compare the measured spectra with simulations.

## VI. Response function simulations

Monte Carlo simulations were performed to reproduce the experimental spectra:

- To simulate the laser photon - relativistic electron scatterings, the laser and the electron beam were modelled as published in [25].
- To simulate the interaction of the so produced γ-ray beams in the collimators and in the LaBr$_3$:Ce detector was simulated using the GEANT4 libraries [1, 7].
- The energy resolution function used was the one of Ref. [10].
- The crystal has been considered bare because, to a first approximation, the response function only depends on the system geometry.

In Fig. 5 the comparison between C2-P55 experimental and simulated spectra is shown at beam energies of 6, 15 and 29 MeV. Because the simulations do not take into account the PMT induced non-linearity effects discussed in the previous section we can deduce a non-linearity curve for each PMT by calculating the differences between the endpoint of the measured and simulated spectra for each beam energy. In Fig. 6 the energy differences are plotted for both PMT's as a function of the beam energy. It results that the energy difference increases with the beam energy, as expected, although the PMT P59 preserves linearity up to almost 20 MeV.

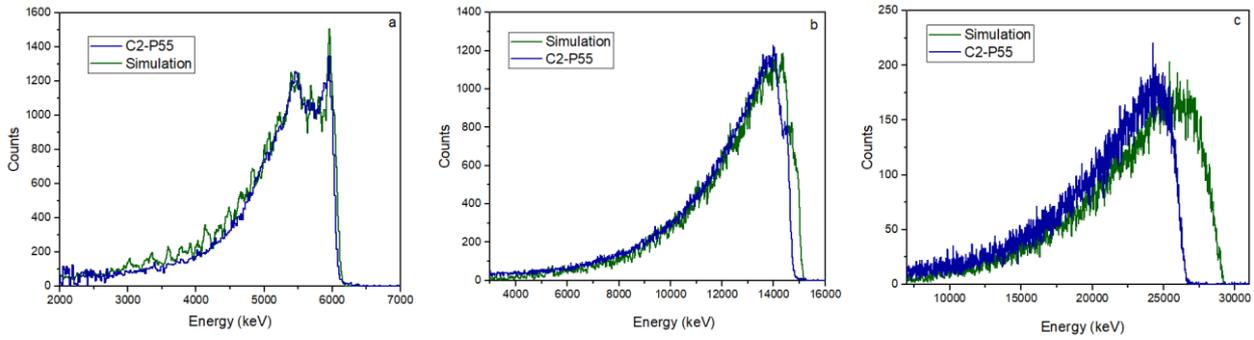

Fig. 5. *The comparison between C2-P55 experimental and simulated spectra at beam energy of 6(a), 15 (b) and 29 MeV (c) is shown. The difference increases as the beam energy increases (color online).*

Interpolating the difference points analytically, we can produce a non-linearity curve for each PMT. The dashed lines in the plot of Fig. 6 show a fit performed using a third order polynomial function on the average values of the two data set recorded with the two crystals. One should remember that the non-linearity curve has been extracted for $E_\gamma > 5$ MeV. These curves represent the non-linearity curve for each PMT. The polynomials of the two curves are also shown in figure 6, the black one for P59 and the red one for P55.

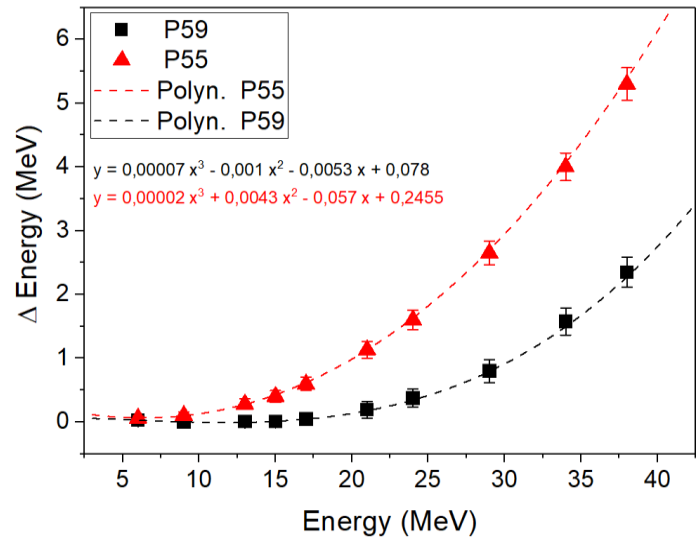

Fig. 6. *The differences between the endpoint of the measured and simulated spectra are plotted as a function of the beam energy. P59 data are indicated by black squares and P55 data by red triangles (color online). Data are fitted by a third order polynomial (indicated by black (P59) and red (P55) dashed lines, respectively) which represent the non-linearity curves for the two PMT's. The polynomials are shown in figure (black (P59) and red (P55)).*

By applying the non-linearity curve to the calibration of the simulated spectra, we obtain spectra that well overlap with the measured ones, as can be seen in Fig. 7, were the comparison is shown for the data recorded with C2-P55 at beam energy of 6, 15 and 29 MeV.

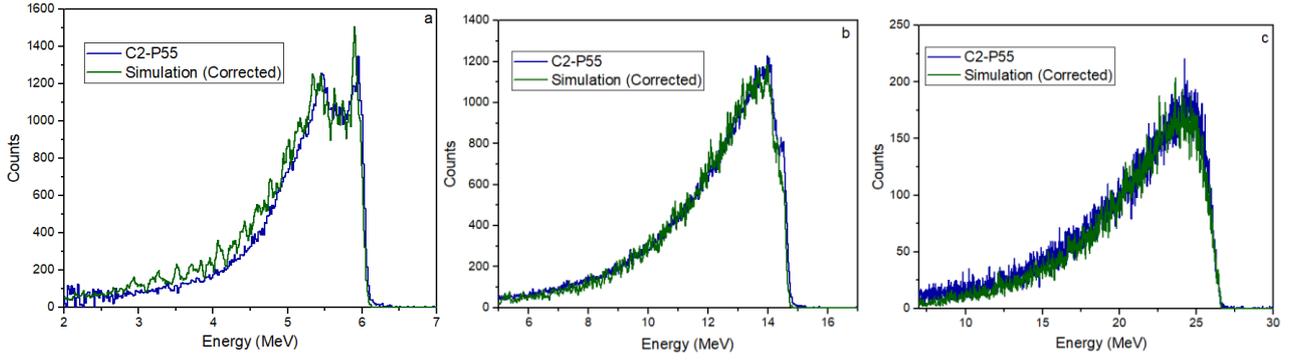

Fig. 7. *The experimental spectra for C2+P55 are compared to the Monte Carlo simulations corrected for the PMT non-linearity using the curve shown in Fig.10 (see text) at beam energy 6 (a), 15(b) and 29 MeV(c).*

Note that both 38 MeV experimental points of Fig. 6 perfectly lie on their fitted non-linearity curves. In fact, as mentioned before, the electronics linearity was tested to be smaller than 0.3% up to 5.5 V output signal, which was the case for all beam energies up to 34 MeV, while for 38 MeV the signal amplitudes were ~5.7 V. This indicates that the PMT non-linearity is much larger than that induced by the used amplifier.

### VII. Interaction and photopeak efficiency from monochromatic beam simulations

In order to obtain information about the full energy peak relative efficiency and the interaction efficiency we performed two simulations in the case of a collimated source and in the case of a non-collimated source, which represent the realistic situation of a nuclear physics experiment, in which the incident γ-rays were emitted isotropically from a point-like source at 20 cm from the detector front face.

The interaction efficiency values, defined as the ratio of number of γ-rays undergoing to at least one interaction in the detector and the number of incident γ-rays, are shown on the left side of Fig. 8 for the collimated and the isotropic beams. In this case, the interaction efficiency increases with beam energy following the trend of the pair production cross section. We observe that in the case of an isotropic source the percentage is lower than in the case of a collimated beam. This behaviour is understandable because the geometrical factor is more important for an isotropic source. All these simulated efficiency values are in agreement with the results of the simulations performed in Ref. [18].

The photopeak efficiencies obtained in the two simulations, defined as the ratio between the number of fully detected γ-rays and the number of incident γ-rays, are shown on the right side of Fig. 8. Although the interaction cross section, mainly due to pair production, increases with energy, we observe that the efficiency decreases with energy. We observe, also in this case, that the efficiency of the realistic isotropic source is lower than in the case of the collimated beam. The geometrical factor for the photopeak efficiency weights crucially compared to the interaction efficiency. The ratio between interaction and photopeak efficiency for the isotropic source, represents a useful estimate of the peak over background ratio for high energy gamma rays in realistic nuclear physics experiment.

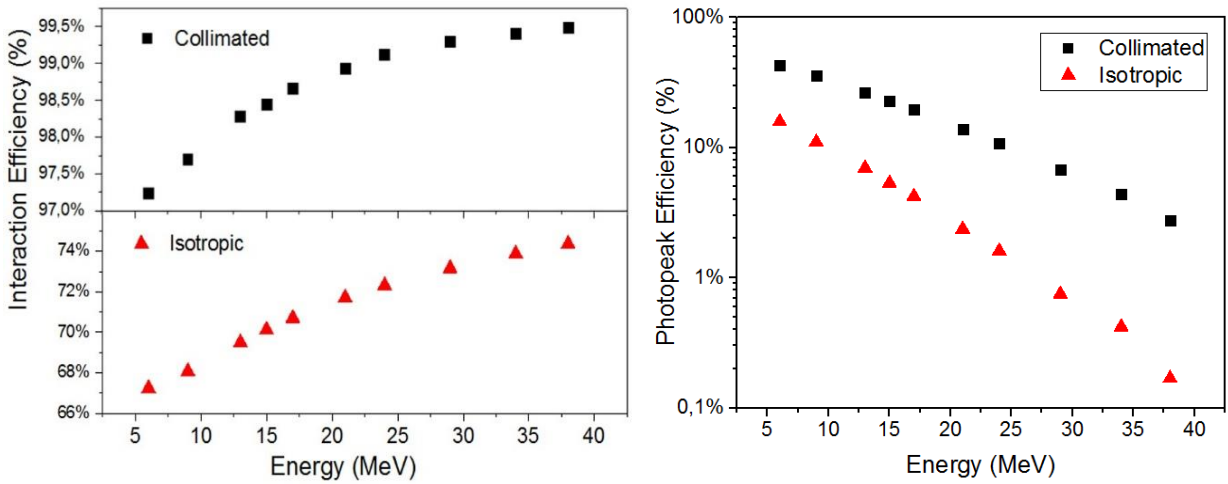

**Fig. 8.** *The trend of interaction efficiency percentage (on the left) and the photopeak efficiency percentage (on the right) is shown. The collimated beam is indicated with filled black squares and the isotropic beam with filled red triangles (color online).*

Conclusions

a. We measured the response function of two scintillator detectors (each consisting of a 3.5″x8″ large volume LaBr$_3$:Ce crystal) to quasi-monochromatic γ-rays with energy ranging from 6 to 38 MeV produced in the NewSUBARU facility [18]. Each crystal was coupled to two different PMT's of the same type, so that four spectra were recorded for each beam energy.

b. We investigated the linear behavior of the crystal, of the PMT together with the VD and the electronics used to handle the signals in a separate way, so that we could identify the origin of possible non-linearity effects. It turns out that:

    - the two crystals respond in the same way to high energy γ-rays and there is no evidence of non-linearity.
    -  The two PMT's suffer from a non-linearity at high energy, which depends on the PMT itself, though the non-linearity curves have a similar trend.

c. We deduced the photopeak efficiency and the interaction efficiency of a 3.5″x8″ LaBr$_3$:Ce crystal for γ-rays energies ranging from 6 to 38 MeV from Monte Carlo simulations both for a collimated incident beam and an isotropic source. The photopeak efficiency turns out to be one order of magnitude smaller in the case of an isotropic source, while the interaction efficiency is about five times smaller.

### Acknowledgement

This work was supported by ENSAR2-PASPAG within the H2020-INFRAIA-2014–2015 Grant Agreement 654002 – ENSAR2-PASPAG. D.F. and I.G. acknowledge the support from the Extreme Light Infrastructure Nuclear Physics (ELI-NP) Phase II, a project co-financed by the Romanian Government and the European Union through the European Regional Development Fund - the Competitiveness Operational Programme (1/07.07.2016, COP, ID 1334). This work was partially supported by INFN Italy, which provided the large-volume LaBr$_3$:Ce scintillators. Thanks to all the collaborators: N. Blasi, F. Camera, B. Million, A. Giaz, O. Wieland, F.M. Rossi, H. Utsunomiya, T. Ari-izumi, D. Takenaka, D. Filipescu, I. Gheorghe.